\documentstyle[epsfig,12pt]{article}

\begin{document}

\begin{center}
{\Large \bf A new method to measure the CP violating phase $\gamma$ using 
$B^{\pm} \to \pi^{\pm}\pi^+\pi^-$ and $B^{\pm} \to K^{\pm}\pi^+\pi^-$ 
decays}

\vskip 1.5 cm
{\large R.E. Blanco, C. G\"obel, 
and R. M\'endez--Galain}
\vskip 1 cm

{\it Instituto de F\'{\i}sica, Facultad de Ingenier\'{\i}a,
Univ. de la Rep\'ublica, \\
CC 30, CP 11000 Montevideo, Uruguay}

\end{center}
\vskip 1.5 cm

\begin{abstract}

A new and simple procedure to measure the angle $\gamma$ 
from $B^{\pm} \to \pi^{\pm}\pi^+\pi^-$ and $B^{\pm} \to K^{\pm}\pi^+\pi^-$ 
decays using SU(3) symmetry is presented. It is based
on a full Dalitz plot analysis of these decays.
All diagrams, including strong and electroweak penguins,
 are considered in the procedure.
 The method is also free from final state interaction problems.
The theoretical error in the extraction of $\gamma$ within the method
 should be of the order of $10^{\rm o}$ or even less. Taking into account
the B-meson production in the first generation of B factories and 
recent measurements from CLEO, this method could bring the best measurement 
of $\gamma$ in the next years.

\end{abstract}

\vfill 

\begin{center}

{\em To appear in Physical Review Letters}

\end{center}

\newpage

For the next years, large accelerator facilities known as B factories are
 going to be running. The goal is to produce a large amount of B mesons 
because their decay should be sensitive to CP violation\cite{babar-intro,gen}. 
As a consequence, one hopes to measure the three Cabibbo-Kobayashi-Maskawa 
(CKM) angles $\alpha\equiv arg({-V_{td}V^*_{tb}/V_{ud}V^*_{ub}}), 
\beta\equiv arg({-V_{cd}V^*_{cb}/V_{td}V^*_{tb}})$ and $\gamma\equiv 
arg({-V_{ud}V^*_{ub}/V_{cd}V^*_{cb}})$ to check the standard model 
 predictions. It is not yet clear which precise measurements will allow 
a clean extraction of $\alpha$ and $\gamma$. Concerning the last 
angle, many interesting methods have been proposed so far but all of 
them suffer from either experimental or theoretical problems. On the 
one side, theoretically clean procedures based in 
$B \to D K$ decays\cite{pap59} suffer from important 
experimental difficulties 
and would demand about 10 years of data taking in order to extract 
$\gamma$ with an error, at least, of the order of 15$^{\rm 
o}$\cite{babar-intro}. 
On the other side, 
other methods based in $B \to \pi\pi$, $B \to \pi K$ or 
$B \to KK$ decays\cite{kpi} have theoretical uncertainties which would imply 
systematic errors in the extraction of $\gamma$ as large as 20$^{\rm o}$
\cite{gronau}. It is then worth looking for new experimental 
procedures to extract $\gamma$ with smaller uncertainties.

Methods proposed so far to measure $\gamma$ are based in the 
study of branching ratio asymmetries in two body decays. In a previous 
letter\cite{prl2} we showed that three body decays could be a more 
interesting tool to extract CP violating phases\cite{helen}. 
The idea is to make use of the fact that Dalitz plot analysis of a 
three body decay gives a direct measurement of {\it amplitudes} instead 
of branching ratios. In other words, one has a direct access to the 
{\it phase} of a given process. In ref. \cite{prl2} we have illustrated 
this general remark applying it to the extraction of $\gamma$ studying the 
decay $B^{\pm} \to \pi^{\pm}\pi^+\pi^-$ . We showed that, due to this 
direct experimental access to 
phases, $\gamma$ could be extracted with a smaller statistical error 
than the usual methods based in two body decays existing in the literature. 
Unfortunately, the method suffers from one difficulty 
also existing in two body decays, i.e., penguin pollution: the bigger is the 
unmeasured penguin contribution compared to the tree one, the 
larger is the systematic error of the method. 

In this letter, making a full use of Dalitz plot analysis features, we
present a first procedure to extract $\gamma$ using three body 
decays which takes penguin contributions 
explicitly into account. 
The method is based in a combined study of two pairs 
of CP conjugated decays $B^{\pm} \to \pi^{\pm}\pi^+\pi^-$ and 
$B^{\pm} \to K^{\pm}\pi^+\pi^-$ \ related by SU(3) symmetry. 
It is free from final state interaction and electroweak penguin problems. 
The theoretical error of the method is due to 
 SU(3) breaking and the uncertainty 
about charm penguin diagrams.  Considering present expectation values
for $\gamma$ and $\beta$, we estimate this error to be of the order of
10$^{\rm o}$ or even less. This value is 
smaller than that 
of other methods experimentally accessible within a few years.

Let us start the description of the method by presenting the main features
of Dalitz plot analysis in our context.
Three body decays of heavy mesons seem to be dominated by intermediate 
resonant decays\cite{pdg}. All these partial intermediate channels, 
together with the direct non-resonant channel interfere to give 
the same --- detected --- three body final state. The Dalitz plot 
analysis is a powerful tool that yields a clear separation of all these 
intermediate channels; moreover, it brings a direct measurement of the 
{\it amplitude}, i.e., magnitude and phase, of the contribution of all the 
intermediate processes. As phases are always measured with respect 
to a given one, in order to extract a weak phase with this method, one needs 
at least two distinct channels with different weak phase \cite{prl2}.
 
This is the case for the two pairs of
decays $B^{\pm} \to \pi^{\pm}\pi^+\pi^-$ and $B^{\pm} \to K^{\pm}\pi^+\pi^-$ 
considered here, thanks 
to the presence of $\chi_{c0}$ as a possible intermediate resonance. 
Indeed, the contribution ${\cal A} (B^\pm \to \chi_{c0} {\cal P^\pm}) 
\times {\cal A} (\chi_{c0} \to \pi^+ \pi^-)$ (where ${\cal P}=\pi$ or $K$) 
has no weak phase while for the other possible channels --- as 
for example those mediated by a $\rho^0$ or $f_0$ resonances --- 
the tree contribution has weak phase $\gamma$. 

Final state interaction problems do not affect this method.
One can easily get rid of them by choosing
 a contribution mediated by an 
isospin 0 resonance\cite{prl2,gerard}
--- such as an $f_0$ or a possible $\sigma$ resonance.
Indeed, a $B^\pm \to f_0{\cal P}^\pm$ decay for example,
 proceeds through a unique isospin amplitude. Thus, 
the method presented in this letter is completely free from final state 
interaction difficulties. In the following, we will present the method using
a $B^\pm \to f_0{\cal P}^\pm$ decay.

Besides tree contributions driven by 
the weak phase $\gamma$, ${\cal A} (B^\pm \to f_0 {\cal P}^\pm)$ has also
penguin
contributions with different weak phases. For ${\cal P}=\pi$, top penguin 
diagrams have weak phase $\beta$ and charm penguin ones have no weak phase;
on the other side, for ${\cal P}=K$, both top and charm penguins have no weak 
phase. Thus, for the four decays, the $f_0$ amplitudes are written as
\begin{equation}
A_1 = T e^{i(\delta_T+\gamma)}+P_te^{i(\delta_{P_t}-\beta)} +  
P_ce^{i\delta_{P_c}} 
\label{1}\end{equation}
\begin{equation}
A_2 = T e^{i(\delta_T-\gamma)}+P_te^{i(\delta_{P_t}+\beta)} + 
P_ce^{i\delta_{P_c}}
\label{2}\end{equation}
\begin{equation}
A_3 = T' e^{i(\delta_{T'}+\gamma)}+P'e^{i\delta_{P'}} 
\label{3}\end{equation}
\begin{equation}
A_4= T' e^{i(\delta_{T'}-\gamma)}+P'e^{i\delta_{P'}}  
\label{4}\end{equation}
where  $A_1={\cal A} (B^+ \to f_0 \pi^+)$,  $A_2={\cal A} (B^- \to f_0 
\pi^-)$, $A_3={\cal A} (B^+ \to f_0 K^+)$ and $A_4={\cal A} 
(B^- \to f_0 K^-)$. In the expressions above, $\delta_T, \delta_{T'}, 
\delta_{P_t}, \delta_{P_c}$ and 
$\delta_{P'}$ are strong (CP conserving) phases, $T$ and $T'$ contain 
both tree and color suppressed contributions, $P_t$ 
includes all strong and electroweak penguin diagrams with weak phase 
$\beta$, $P_c$ includes all those
with no weak phase and $P'$ includes all penguins. In other words,
Eqs. (\ref{1}--\ref{4}) include {\it all} kind of diagrams 
contributing to these decays\cite{fn2}. Some of these decays can be related by 
SU(3) symmetry, as we will see below.

The left-hand sides in Eqs. (\ref{1}--\ref{4}), i.e. $A_i, (i=1, ..., 4)$, are 
{\it directly measured} complex numbers, that is, 8 real 
independent quantities. This is one of the main claims of this letter: 
Dalitz plot analysis brings more independent measurements than usual two body 
branching ratios; one thus has more information available for each decay 
that can be used to treat penguin and other pollutions.

As a first step, we will exclude from our analysis charm penguin contributions
which appear only in the first two equations. Note that this is not a 
strong assumption.
Indeed, in $B^{\pm} \to \pi^{\pm}\pi^+\pi^-$ decays, as in 
$B \to \pi\pi$, one expects penguin 
contributions to be of the order of 20\% of tree ones\cite{gen}. 
Moreover, charm penguin amplitudes are expected to be not larger than 
half of top ones\cite{Pc}. Thus, in order to resolve in $\gamma$ in Eqs. 
(\ref{1}--\ref{4}),
the influence of charm penguin is expected to be small. In spite of this, in 
the last part of this letter, we will explicitly study their influence 
in our results.

Now, we make use of SU(3) flavor symmetry.
All contributions included in $T'$ are identical to those in $T$ 
except for an $s$ 
quark replacing a light one. (Note that one cannot make the same
assumption for the penguin sector because in $P_t$ there are two different 
topologies while in $P'$ there is only one.)
One can then write,
\begin{equation}
T'=\lambda T
\label{SU3}
\end{equation}
\begin{equation}
\delta_T'=\delta_T
\label{SU3d}\end{equation}
where $\lambda = V_{us}/V_{ud} \approx 0.2.$

The validity of these assumptions will be discussed later in this letter.

One then has a system with 8 real equations and 8 
unknown quantities. After a simple algebra to eliminate  
unwanted quantities, one can reduce the set of 8 equations 
to a simpler system containing the desired variable $\gamma$:
\begin{equation}
\frac{e^{i(\delta_T-\gamma)} - e^{i(\delta_T+\gamma+2\beta)}}{|e^{i(\delta_T
+\gamma)}-e^{i(\delta_T-\gamma)}|}
= \frac{\lambda \; (A_2 - A_1e^{i2\beta})} {|A_3-A_4|} 
\label{3x3-1}\end{equation}
\begin{equation}
arg(A_3-A_4) = arg(e^{i(\delta_T+\gamma)}-e^{i(\delta_T-\gamma)}) .
\label{3x3-2}\end{equation}

This system, containing 3 real equations, 
allows us to obtain $\gamma$, $\beta$ and $\delta_T$. 
At this stage, one could expect this method to provide not only a 
measurement of $\gamma$ but also an independent measurement of $\beta$. 
Unfortunately, this is only true if Eqs. (\ref{3x3-1}) and (\ref{3x3-2}) 
were exact. 
This is not the case due to the theoretical assumptions made above. 
The question is how does any error in these 
equations propagate to the actual extraction of $\gamma$ and $\beta$. 

We found that $\beta$ is too sensitive to small uncertainties in 
these equations. This is simply due to the fact that $\beta$ is only present
in in Eqs. (\ref{1}) and (\ref{2}), in a term which is small. Thus, 
small uncertainties in coefficients  $A_{1,2}$ imply large
uncertainties in $\beta$.  

As a consequence, the method is
not well suited to providing an independent measurement of $\beta$. Fortunately, 
in the next years, the phase $\beta$ will be known within an error of a
few degrees. Our strategy is thus to consider $\beta$ as a known parameter.
We then have two independent variables,
$\gamma$ and $\delta_T$ with only two equations, e.g., real 
and imaginary parts of Eq. (\ref{3x3-1}) \cite{foot4}.
 
Let us now study the theoretical errors of the method and their
influence in the extraction of $\gamma$.
The sources of systematic errors are 1) the validity of SU(3) 
symmetry assumptions and 2) charm penguins. 

SU(3) symmetry is only assumed for those terms carrying the weak phase 
$\gamma$, i.e., diagrams included in $T$. This point represents another 
important issue of this method: one does not need to make SU(3) assumptions
about penguin contributions.
The two main contributions to $T$ 
are tree and color suppressed diagrams. Factorization corrections 
to exact SU(3) 
symmetry in both diagrams transform Eq. (\ref{SU3}) in
\begin{equation}
T'=\frac{f_K}{f_{\pi}} \lambda T
\label{su3-2}\end{equation}
where $f_K$ and $f_{\pi}$ are the kaon and pion decay constants, 
respectively. As $f_K/f_\pi \approx 1.2$, this represents a 20\% correction 
with respect to the assumption of exact SU(3) symmetry of Eq. (\ref{SU3}). 
The importance of 
non-factorizable corrections to Eq. (\ref{su3-2}) is not yet 
clear but there are usually assumed to be small. Both theoretical
\cite{su3-gronau} and experimental\cite{ref-su3-2} educated studies seem 
to favor non-factorizable corrections not larger than 10\%. 

We have studied numerically the influence of this uncertainty 
when solving Eqs. (\ref{3x3-1}). 
For this, after changing (\ref{SU3}) to (\ref{su3-2}) we have assumed
an extra 10\% theoretical uncertainty in the relation between $T'$ and $T$ and 
studied the propagation of this error in the extraction of $\gamma$. 
More precisely, we have first assumed a given set of 
values for the various parameters entering the right-hand-sides of 
Eqs. (\ref{1}--\ref{4}) --- including
$\gamma$. With them, we have {\it calculated} 
the quantities $A_i, (i = 1, ..., 4)$, using Eqs. 
(\ref{1}--\ref{4}) considering explicitly a 10\% correction to 
(\ref{su3-2}). Then we have solved 
the system of Eqs. (\ref{3x3-1}) --- 
which assumes that there is no such a correction --- to find $\gamma$. We 
have finally compared the latter with the originally assumed value for 
$\gamma$.

We found that the amount of this error in extracting $\gamma$
only depends on the actual value of $\gamma$ and $\beta$, as shown in
Fig. 1. This error was found to be
 independent of the actual values of $T, P_t, P', \delta_T, 
\delta_{P_t}$ and $\delta_{P'}$. 
Fig. 1 shows that the smaller $\beta$, the larger the errors. When 
$\beta \to 0$, the error diverges. 
This is simply because, when $\beta$ is exactly zero, Eqs. (\ref{1}--\ref{4})
 with $P_c=0$, are no longer independent; thus Eqs. (\ref{3x3-1}) only 
admit solutions when SU(3) symmetry assumption is exact.

\begin{figure}
\begin{center}
\mbox{\epsfig{file=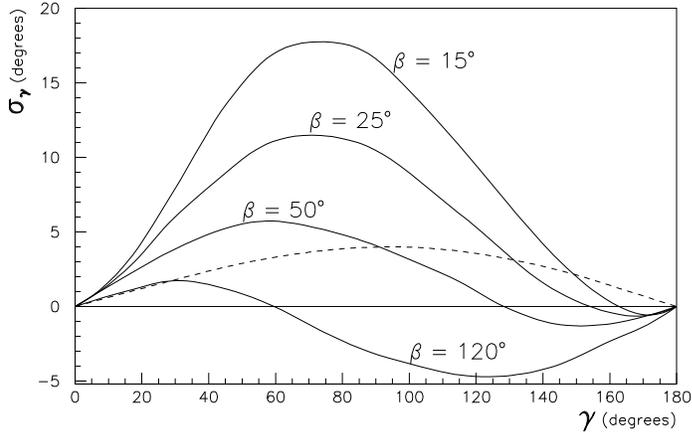,width=10cm}}
\caption{The error in the extraction of $\gamma$ as a function of $\gamma$.
Solid curves are the error due to the uncertainty 
in the SU(3) assumption, for 4 different values of $\beta$. The dashed curve
is the error due to charm penguins.}
\end{center}
\label{F1}
\end{figure}

The other assumption related to the SU(3) symmetry, i.e., Eq. (\ref{SU3d}), 
has no influence in the result. To find this, we proceeded in the
 same way as described above ---
this time assuming $\delta'_T \neq \delta_T$ when calculating $A_i$.
We found that the error in finding $\gamma$ due to this assumption 
is always negligible, independently of the values 
of $T, P_t, P', \delta_T, \delta_{P_t}, \delta_{P'}, 
\beta$ and $\gamma$.

Let us now consider the error due to charm penguins. They would only affect 
 Eqs. (\ref{1}) and (\ref{2}) since in the $B^{\pm} \to K^{\pm}\pi^+\pi^-$ 
 decays charm penguins
have the same weak phase as top ones and are thus already included in $P'$. 
This is a crucial remark, because in the $B^{\pm} \to \pi^{\pm}\pi^+\pi^-$ 
decay $P_t/T$ is expected 
to be small\cite{gen}; 
 thus, as $\gamma$ is present in the {\it T}-term, any error in the penguin 
sector is less important. The actual importance of these neglected terms 
is in fact unknown at present, but a crude estimate gives\cite{Pc} $0.2 
\leq P_c/P_t \leq 0.5$.

The theoretical uncertainty in the value of this ratio propagates 
to the actual extraction of $\gamma$. To estimate the amount of this error, 
we made the same kind of numerical analysis as described above to the 
study of SU(3) symmetry assumptions. In other words, we have calculated
the quantities $A_i$ assuming that $P_c$ is not zero, and then solve
Eqs. (\ref{3x3-1}) --- where $P_c$ is not present.
We found that in this case, the error in the extraction 
of $\gamma$ does not depend on the actual value of $\beta$, but it does
depend on the value of $\gamma$, as shown in the dashed curve of Fig. 1. 
As in the SU(3) case, these results are independent of the actual 
values of $\delta_T, \delta_P, \delta'_P$ and 
$P'$. Nevertheless, the dashed curve in  Fig. 1 does depend on the actual value of
$P_t/T$. More precisely, the error is proportional to the value of this ratio. 
Results presented in Fig. 1 have been obtained using the expected 
values of $P_t/T = 0.2$ \cite{foot2}, $P_c/P_t = 0.35$ \cite{Pc} and 
$\delta_{P_c} = \delta_{P_t}$ \cite{dighe-15}. These results 
do not change if the last
assumption is released, and change little if $P_c/P_t$ takes another value
within the range 0.2 -- 0.5.

Finally, we have studied how an uncertainty in $\beta$ propagates to the 
extraction of $\gamma$. Assuming $\beta$ is known with an error of 
3$^{\rm o}$ -- which is the expected error after 4 years of running of 
BaBar and Belle -- this contributes with less than 1$^{\rm o}$ in the error
to extract $\gamma$.

Let us now analyse the total theoretical error of the method. 
Adding in quadrature all the errors we have discussed above,
we conclude the following. If $\beta$ is 50$^{\rm o}$ or more, 
the total error of the method is always small: 5$^{\rm o}$  or less.
If $\beta$ turns out to have the central value estimated by
Standard Model predictions\cite{ali-london}, i.e. 25$^{\rm o}$
(which is also in reasonable agreement with recent reports from Belle and 
BaBar\cite{beta}), then the error may be larger: if 
$\gamma$ is in the range 50$^{\rm o}$ -- 80$^{\rm o}$, as dictated by 
an overall analysis of the unitary triangle\cite{ali-london}, then 
the error is of the order of 11$^{\rm o}$; if $\gamma$ is in the range
 100$^{\rm o}$ --
130$^{\rm o}$, as suggested by experimental constraints based in 
non-leptonic B decays\cite{cleo-gamma}, then the error in 
extracting $\gamma$ is of the order of 5$^{\rm o}$ to 8$^{\rm o}$.

All these values for the theoretical error are smaller that what one 
expects from 
other methods to extract $\gamma$, proposed so far. This is valid as 
far as the value of $\beta$ is larger than 10$^{\rm o}$. 

It is important to stress that the method is also self contained 
for the determination of its systematic error. Indeed, once the 
actual values of $\gamma$ and $P_t/T$ are 
determined from the experiment, the systematic error can easily be 
read from the numerical study made above. 

Our last comment deals with the experimental feasibility of this procedure. 
The channels involved in the analysis only require the measurement 
of charged pions and kaons and do not require the necessity of 
tagging (since the method only deals with charged B meson decays).
In ref. \cite{prl2} we made a simulation of $B^{\pm} \to \pi^{\pm}\pi^+\pi^-$ 
decays; we
showed that the extraction of partial 
amplitudes -- i.e., what we call $A_i, (i=1, ..., 4)$ in this letter -- 
could be done, with relatively small statistical errors, with 
about 1000 reconstructed events in each Dalitz plot. 
Recent CLEO data\cite{cleo-rec} found that 
$B^\pm \to 
{\cal P}^\pm \rho^0$ (with ${\cal P} = K, \pi$) have branching ratios 
of the order of $10^{-5}$. 
Thus, assuming a total branching ratio of the 
decays $B^{\pm} \to \pi^{\pm}\pi^+\pi^-$ and $B^{\pm} \to K^{\pm}\pi^+\pi^-$ 
of the order of 
$2 \times 10^{-5}$ and a 60\% 
reconstruction efficiency \cite{babar-intro}, this method may provide a 
good measurement of $\gamma$ after about 4 years of running of the 
first generation of B machines.

In summary, we have presented in this letter a method to measure 
the weak CKM angle $\gamma$ which deals with all strong
and electroweak penguin amplitudes. 
It exploits a general procedure to extract 
weak angles using Dalitz plot analysis in three body B meson decays. 
Here, we discuss a combined study of two pairs of CP conjugated 
decays, $B^{\pm} \to \pi^{\pm}\pi^+\pi^-$ and $B^{\pm} \to K^{\pm}\pi^+\pi^-$ 
$\;$ which allows the 
treatment of penguin amplitudes.
 Moreover, tree 
and penguin magnitudes and strong phases are obtained from the experimental 
procedure. The method brings a measurement of $\gamma$ free from 
final state interaction problems. 

The procedure has two theoretical sources of uncertainty. First, it is based 
in SU(3) symmetry assumptions, even though only for the tree sector, 
not in the penguin one.  
Second, charm penguins can only be included in the method by considering 
a model to estimate $P_c/P_t$. According to the 
present knowledge, charm penguins would introduce a small 
systematic uncertainty. Combining both sources of error, the total 
theoretical error of the method presented in this letter is not larger 
than $5^{\rm o}$ if $\beta$ is larger than 50$^{\rm o}$; if $\beta$ has
the expected value of 25$^{\rm o}$, then the error is of the order of
5$^{\rm o}$ -- 8$^{\rm o}$ if $\gamma$ turns out to be in the
second quadrant and of the order of 11$^{\rm o}$ if it is in the first 
one. Thus, one expects the 
theoretical error of this method to be smaller than that of other 
procedures proposed so far to extract $\gamma$.

The method presented here should be experimentally feasible at BaBar 
and KEK. It could then provide the best knowledge of $\gamma$ in the 
next years. In any case, this method is based on a new idea and a different
experimental procedure, if compared to usual methods based in two 
body decays. It will then bring an independent measurement of $\gamma$.


\begin{thebibliography}{999}

\bibitem{babar-intro} P.F. Harrison, ed. and H.R. Quinn, ed., 
{\it The BaBar Physics Book} (SLAC-R-504, 1998).

\bibitem{gen} For a general review see: Y. Nir and H.R. Quinn, in 
{\it B decays}, edited by S. Stone (World Scientific, Singapore, 1994), 
p. 520; M. Gronau, Nucl. Phys. (Proc. Suppl.) {\bf B38}, 136 (1995).


\bibitem{pap59} M. Gronau and D. Wyler, Phys. Lett. B {\bf 265}, 172 
(1991); I. Dunietz, Phys. Lett. B {\bf 270}, 75 (1991); D. Atwood, 
I. Dunietz, and A. Sony, Phys. Rev. Lett. {\bf 78}, 3257 (1997).

\bibitem{kpi} For a review see for example R. Fleischer, hep-ph/9904313 and 
references therein.

\bibitem{gronau} M. Gronau, hep-ph/0001317; and references therein.

\bibitem{prl2} I. Bediaga, R.E. Blanco, C. G\"obel, and R. M\'endez-Galain, 
Phys. Rev. Lett. {\bf 81}, 4067 (1998).

\bibitem{helen} The use of Dalitz plot analysis to measure CP violating angles
has first been introduced, in a quite 
different context, in: G. Burdman and J.F. Donoghue, Phys. Rev. D
{\bf 45} 187 (1992) and A.E. Snyder and H.R. Quinn, Phys. Rev. D
{\bf 48}, 2139 (1993).


\bibitem{pdg} Particle Data Group, C. Caso {\it et al.}, Eur. Phys. J. 
{\bf C3}, 1 (1998).

\bibitem{gerard} J.M. Gerard and J. Weyers, Eur. Phys. J. 
{\bf C7}, 1 (1999).

\bibitem{fn2} Only penguins with a quark up in the loop are 
excluded from the analysis of the $B^{\pm} \to \pi^{\pm}\pi^+\pi^-$ decay. 
Nevertheless, they are 
expected to be even smaller than charm penguins.

\bibitem{Pc} A.J. Buras and R. Fleischer, Phys. Lett. B {\bf 341}, 379 (1995).

\bibitem{foot4} Note that, in fact, Eq. (\ref{3x3-2}) is 
independent of $\gamma$: its right-hand side is
$arg(2 i\; e^{i\delta_T}\; \sin\gamma ) = \delta_T + \pi/2$, when $\gamma$
varies between 0$^{\rm o}$ and 180$^{\rm o}$.

\bibitem{su3-gronau} M. Gronau and D. Pirjol, Phys. Rev. D {\bf 62}, 077301 
(2000).

\bibitem{ref-su3-2} M. Gronau, O.F. Hernandez, D. London, and J.L. Rosner, 
Phys. Rev. D {\bf 52}, 6374 (1995); M. Athanas et al., CLEO Collaboration, 
Phys. Rev. Lett. {\bf 80}, 5493 (1998).

\bibitem{foot2} In fact, present estimates correspond to the 
ratio penguin-to-tree diagrams while the {\it T}-term also has color 
suppressed diagrams. Nevertheless, as tree contributions are expected to 
dominate, this ratio should not differ significantly.

\bibitem{dighe-15} A. Dighe, hep-ph/9811506 and references therein.

\bibitem{ali-london} A. Ali and D. London, Eur. Phys. {\bf C9} 687 (1999).

\bibitem{beta} B. Aubert {\it et al.}, BaBar Collaboration, hep-ex/0008048;
 H. Aihara {\it et al.}, Belle Collaboration, hep-ex/0010008.

\bibitem{cleo-gamma} Y. Kwon et al., CLEO Collaboration, hep-ex/9908039.

\bibitem{cleo-rec} C.P. Jessop et al, CLEO Collaboration, Phys. Rev. Lett. 
{\bf 85}, 
2881 (2000).


\end{thebibliography}
\end{document}